\documentclass{edp-conf}
\usepackage{graphicx}
\begin{document}

\TitreGlobal{SF2A 2004}


\title{Earth Orientation and Temporal Variations of the Gravity Field}

\author{Bourda, G.}\address{Observatoire de Paris, SYRTE/UMR 8630-CNRS, \\ 
61 avenue de l'Observatoire, 75014 Paris, France, 
email: bourda@syrte.obspm.fr}

\runningtitle{Earth Rotation and Variable Gravity Field}

\setcounter{page}{237}
\index{Author1, A.}
\index{Author2, B.}
\index{Author3, C.}

\maketitle

\begin{abstract} 
The determination of the Earth gravity field from space geodetic techniques now allows us to obtain the temporal variations of the low degree coefficients of the geopotential, combining the orbitography of several satellites (e.g. Lageos1, Lageos2, Starlette). These temporal variations of the Earth gravity field can be related to the Earth Orientation Parameters (EOP) through the inertia tensor. 
This paper shows these relations and discusses how such geodetic data can contribute to the understanding of the variations in EOP. 
\end{abstract}


\section{Introduction}

The Earth Orientation is generally considered as (i) Earth rotation axis movements in space (precession-nutation), (ii) Earth rotation axis movements in the Earth (polar motion), or (iii) Earth rotation speed variations (exces in the length of the day). These movements come from Earth inside masses distributions.

The Earth gravity field can give us information about this distribution of masses because nowadays we can determine the variations of the Earth gravity field by space geodetic techniques. 

Hence, there is a link between the variations of the Earth gravity field and the variations of the Earth Orientation Parameters. And the high accuracy now reached in the VLBI (Very Long Baseline Interferometry) Earth Orientation Parameters (EOP) determination requires looking further at the various geophysical contributions to variations in EOP. So we investigate here if this variable gravity field can be valuable for the improving the modelisation of the Earth rotation.

\section{Theory}

  \subsection{Link between Earth Orientation and Inertia}

The fundamental equations for the rotation of the Earth in an inertial frame are Euler's dynamical equations, based on the conservation of the angular momentum $\vec H$ of the Earth under an external torque $\vec L$ (Lambeck 1980):
\begin{equation}\label{eq:Euler}
\dot{\vec H} = \vec L  
\end{equation}
For a non-rigid Earth, these equations in a rotating frame become:
\begin{equation}\label{eq:Euler_non_rigid}
\frac{\mbox{d}}{\mbox{d}t}\left[ I(t)~\vec \omega + \vec h(t)\right] + \vec \omega \wedge \left[ I(t)~\vec \omega + \vec h(t)\right] = \vec L
\end{equation}
where the inertia tensor $I$ is time dependent, as well as the relative angular momentum $\vec h$ ,and $\vec \omega$ is the Earth instantaneous rotation vector which direction is the one of the rotation axis and which norm is the rotation speed. It depends on the Earth Orientation Parameters (EOP). The Inertia Tensor, which is symetric, can be written as:
\begin{equation}
I = 
\left[
\begin{array}{ccc}
I_{11}  &  I_{12}  &  I_{13}   \\
I_{12}  &  I_{22}  &  I_{23}   \\
I_{13}  &  I_{23}  &  I_{33}
\end{array}
\right]
= 
\left[
\begin{array}{ccc}
A+c_{11}  &  c_{12}    &  c_{13}   \\
c_{12}    &  B+c_{22}  &  c_{23}   \\
c_{13}    &  c_{23}    &  C+c_{33}
\end{array}
\right]
\end{equation}
with $(A,B,C)$ the constant part and $c_{ij}$ ($i=1,2,3$) the variable part of the Inertia Tensor.

  \subsection{Link between Inertia and Earth Gravity Field}

The Earth gravity field of the Earth devived from the external gravitational potential $U$ which is expressed in a spherical harmonic expansion as (Lambeck 1980):
\begin{equation}
U(r,\phi,\lambda) = \frac{GM}{r} \biggl[ 1 + \displaystyle \sum_{n=2}^{\infty}~ \sum_{m=0}^{n} ~\biggl( \frac{R_e}{r} \biggr)^n  ~(C_{nm}~\cos(m\lambda) + S_{nm}~\sin(m\lambda)) ~P_{nm}(\sin~\phi)  \biggr] 
\end{equation}
where $r$ is the geocentric distance, $\phi$ the latitude and $\lambda$ the longitude of the point at which $U$ is detremined. $G$ is the gravitational constant, $M$ and $R_e$ are the mass and the equatorial radius of the Earth, respectively. $C_{nm}$ and $S_{nm}$ are the Stokes coefficients of degree $n$ and order $m$, and $P_{nm}(\sin \phi)$ are the Legendre polynomials. Hence the second-degree Stokes coefficients can be directely related to the Inertia tensor components (Lambeck, 1988):
\begin{eqnarray}\label{eq:Coeffs_C_Inertie}
C_{20} & = & - \frac{I_{33} - \frac{1}{2} (I_{11}+I_{22})}{M ~{R_e}^2}            \nonumber  \\
C_{21} & = & - \frac{I_{13}}{M ~{R_e}^2}  ~~~~   C_{22} = \frac{I_{22}-I_{11}}{4 ~M~{R_e}^2} \\  
S_{21} & = & - \frac{I_{23}}{M ~{R_e}^2}  ~~~~   S_{22} = - \frac{I_{12}}{2 ~M~{R_e}^2} \nonumber
\end{eqnarray}

\section{Practical links}

We just have shown that the Earth rotation (with $\vec \omega$ and the EOP) could be related to the Earth gravity field (with the degree 2 Stokes coefficients). Then, we investigate now how we can link each EOP with these coefficients.

  \subsection{Earth Rotation speed}
  
\begin{figure}[h]
\begin{minipage}[c]{.46\linewidth}
\centering
\includegraphics[width=4cm, angle=-90]{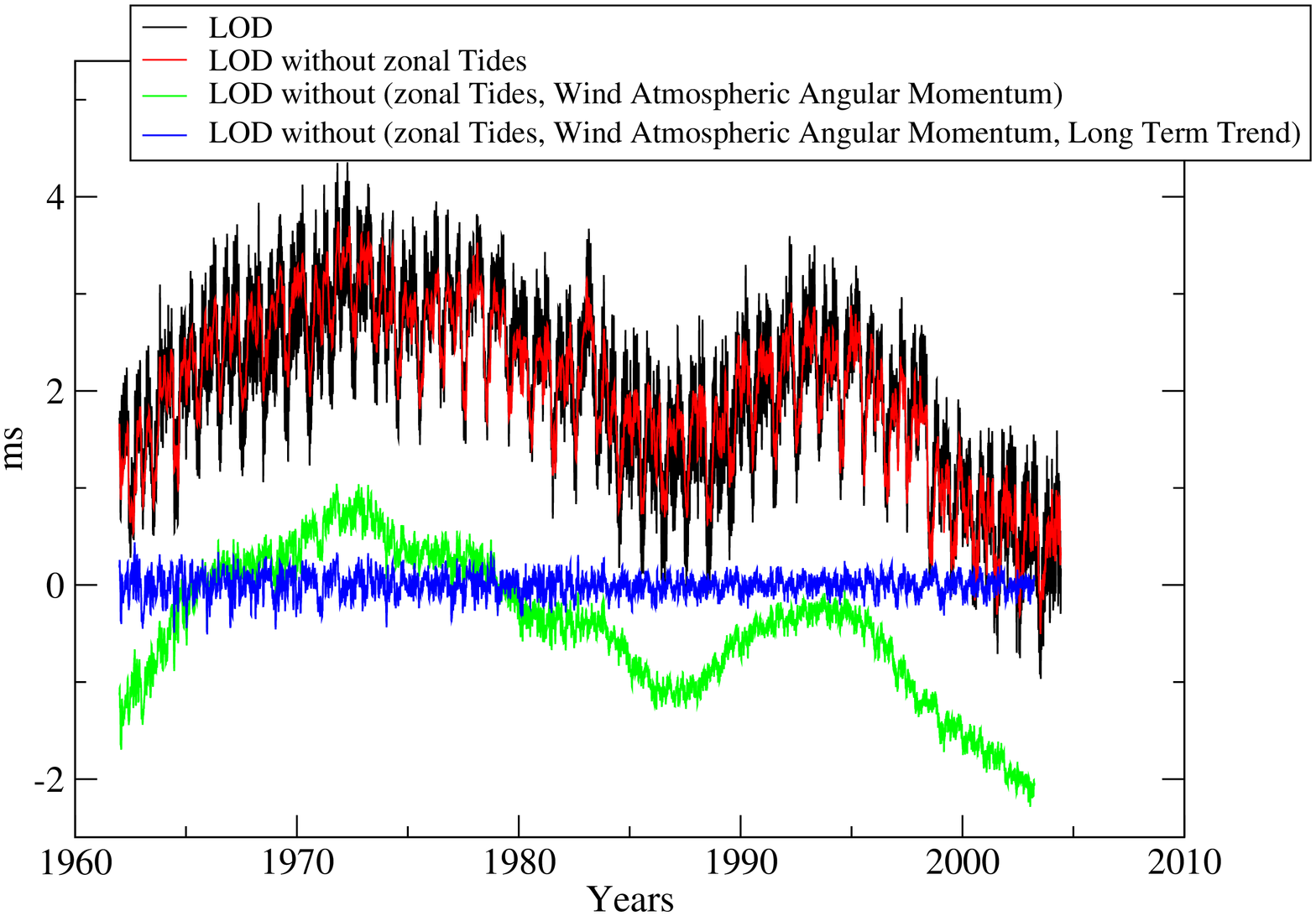}
\caption{Exces in the length of the day: various components of this $\Delta(LOD)$ are removed.}
\label{fig:Bourda_fig1}
\end{minipage}
\hfill  
\begin{minipage}[c]{.46\linewidth}
\centering
\includegraphics[width=4cm, angle=-90]{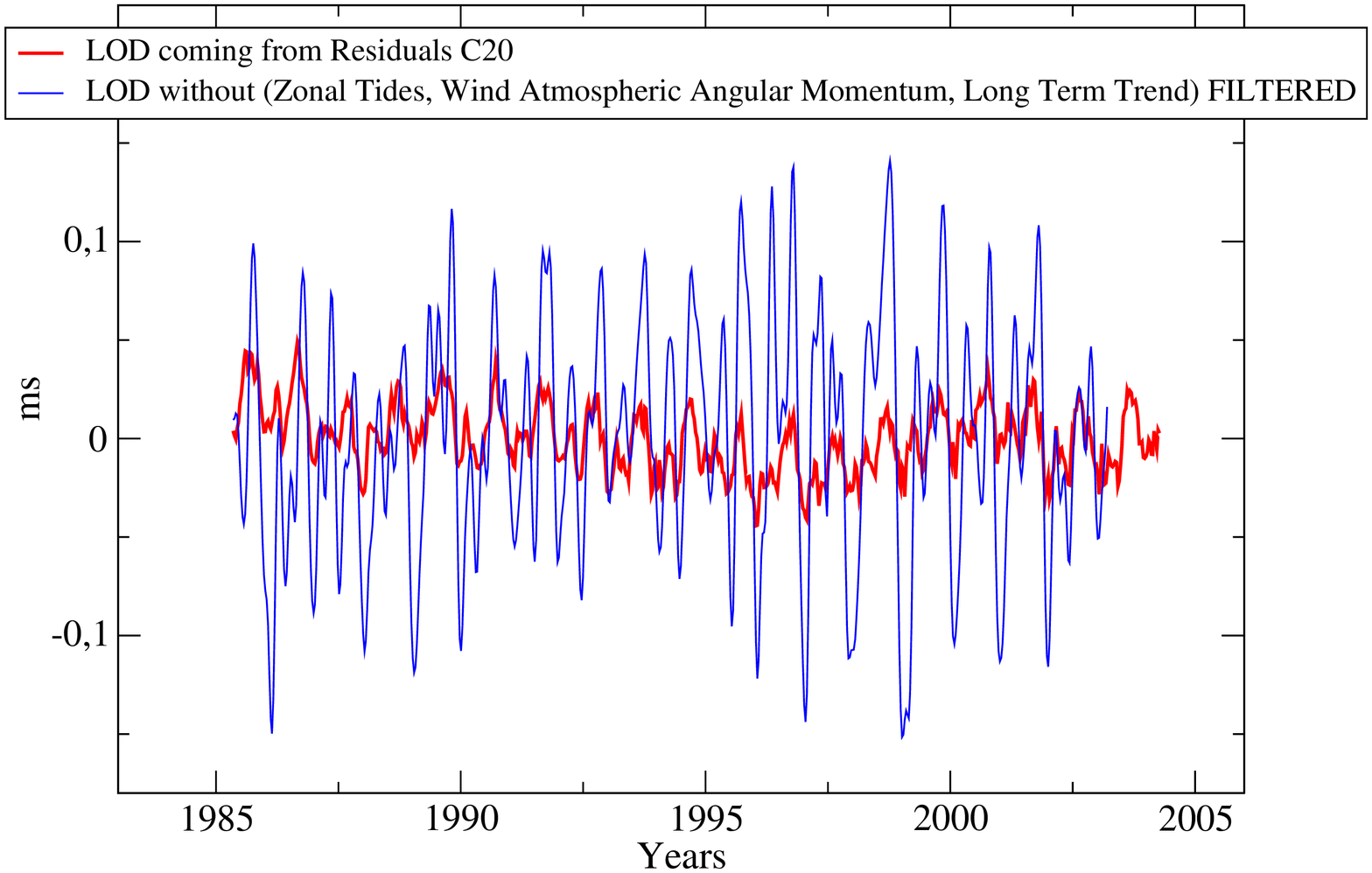}
\caption{Exces in the length of the day obtained with $\Delta C_{20}$ data and compared with the classical one corrected from other effects.}
\label{fig:Bourda_fig2}
\end{minipage}
\end{figure}

The exces in the length of the day $\Delta(LOD)$ (with respect to a mean LOD) can be related to (i) the third component $c_{33}$ of the variable part of the Inertia tensor and (ii) the third component $h_3$ of the relative angular momentum of the Earth, ignoring the external torques:
\begin{equation}\label{eq:LOD}
\frac{\Delta(LOD)}{LOD_{mean}} = \frac{c_{33}}{C} + \frac{h_3}{C~\Omega}
\end{equation}
Moreover, with the help of Eq.~(\ref{eq:Coeffs_C_Inertie}), we can write:
\begin{equation}\label{eq:c33}
c_{33} (t) = \frac{1}{3}~\Delta Tr(I) - \frac{2}{3} ~M~{R_e}^2 ~\Delta C_{20} (t)
\end{equation}
where $\Delta Tr(I)$ is the variation in time of the sum of the diagonal elements of the Inertia tensor. We can consider that it is equal to zero (Rochester \& Smylie 1974). Then, we can obtain:
\begin{eqnarray}\label{eq:LOD_fin}
\frac{\Delta(LOD)}{LOD_{mean}} & = &  - 0.7 ~\frac{2}{3 ~C_m} ~M~{R_e}^2 ~\Delta C_{20} +  \frac{h_3}{C_m~\Omega}  
\end{eqnarray}
where the coefficient $0.7$ accounts for the loading effects and $C_m$ is the third moment of inertia of the Earth's mantle (Barnes et al. 1983). Then we have compared the $\Delta(LOD)$ obtained with Eq.~(\ref{eq:LOD_fin}) and the $\Delta C_{20}$ data in Fig.~\ref{fig:Bourda_fig2} with the one usually used but corrected from zonal tides, atmopheric wind effects ($h_3$) and long terms (see Fig.~\ref{fig:Bourda_fig1}).

  \subsection{Precession-Nutation}
  
The study of the Earth precession nutation angles variations influenced by the temporal variations of the $C_{20}$ coefficients of the geopotential is developped in the article of Bourda \& Capitaine (2004). It is based on the works of Williams (1994) and Capitaine et al. (2003) which considered secular terms for the $C_{20}$ variations, whereas we consider also annual and semi-annual ones.

  \subsection{Polar Motion}

The polar motion $p = x_p - i~y_p$, where $x_p$ and $y_p$ are the components of the rotation axis in space can be theoretically related to the degree 2 and order 1 coefficients of the Earth gravity field:
\begin{eqnarray}\label{eq:p}
p + i~\frac{\dot{p}}{\sigma_r} & = & \frac{1}{\Omega~(C-A)} (\Omega~c+h)  \nonumber \\
                               & = & \frac{1}{\Omega~(C-A)} \left( -M~{R_e}^2~\Omega ~(\Delta C_{21}+i~\Delta S_{21}) + h \right)  
\end{eqnarray}
where $c = c_{13}+i~c_{23}$ is related to $\Delta C_{21}$ and $\Delta S_{21}$ with Eq.~(\ref{eq:Coeffs_C_Inertie}), and $h=h_1+i~h_3$.

\section{Conclusions}

The part of the length of the day obtained with the $\Delta C_{20}$ data corresponds to gravitational terms. Then we have compared $\Delta(LOD)$ corrected from the movements terms (as atmospheric ones), the zonal tides and the decadal terms (from magnetic effects in the core-mantle boundary). But the residual term has an amplitude of the order of $50$ $\mu$s, whereas the better precision on these LOD data is of the order of $10$ $\mu$s. 

The effect of the variable gravity field on the polar motion can be investigated now, using Eq.~(\ref{eq:p}). 

Finally, we find a 18.6-yr periodical effect on the precession angle development in longitude with a sinus term of about $105$ $\mu$as (Bourda \& Capitaine 2004). 

In the future, the static gravity field and its temporal variations coming from the GRACE satellite will be very usefull for these kind of studies, because they are very precise.


\end{document}